\documentclass[superscriptaddress,preprint,prb,showpacs,preprintnumbers,amsmath,amssymb]{revtex4}

\usepackage [dvips]{epsfig}
\usepackage {graphics}
\usepackage {graphicx}
\usepackage{graphicx}
\usepackage{dcolumn}
\usepackage{bm}

\begin{document}

\title{Electron localization : band-by-band decomposition, and application to oxides.}

\author{M. Veithen}
\affiliation{D\'epartement de Physique, Universit\'e de Li\`ege,
B-5, B-4000 Sart-Tilman, Belgium}

\author{X. Gonze}
\affiliation{Unit\'e de Physico-Chimie et de Physique des Mat\'eriaux (PCPM),
      place Croix du Sud, 1, B-1348 Louvain-la-Neuve, Belgium}

\author{Ph. Ghosez}
\affiliation{D\'epartement de Physique, Universit\'e de Li\`ege,
B-5, B-4000 Sart-Tilman, Belgium}

\date{\today}

\begin{abstract}
 
Using a plane wave pseudopotential approach to density functional
theory we investigate the electron localization length in various oxides. 
For this purpose, we first set up a theory of the
band-by-band decomposition of this quantity,  
more complex than the decomposition of the 
spontaneous polarization (a related concept), because of the
interband coupling.  We show its 
interpretation in terms of Wannier functions 
and clarify the effect of the pseudopotential
approximation. 
We treat the case of different oxides: BaO, $\alpha$-PbO, BaTiO$_3$
and PbTiO$_3$. We also investigate the 
variation of the localization tensor during the ferroelectric
phase transitions of BaTiO$_3$ as well as its
relationship with the
Born effective charges.

\end{abstract}

\maketitle

\section{Introduction}

In the study of periodic crystalline solids, the electronic
ground-state wave function is usually described in
terms of Bloch functions, delocalized on the whole system. As
a consequence, for a long time,
the understanding of electron localization in
crystalline solids was mainly based on
approximate pictures. Nevertheless, the basics 
of a quantitative characterization of electron
localization had already been formulated by
W. Kohn~\cite{kohn_loc} in 1964.
Recently this problematic was renewed, 
thanks to the development of the theory of
polarization based on
Berry phases~\cite{rdks,vb,resta94,ortiz_berry_prb}, 
and the rigorous definition of the position operator in
periodic systems~\cite{resta_loc_prl,ressor_loc_prl,aligia_loc_prl,resta_rev}.
These ideas have been
further developped using a
cumulant generating function approach~\cite{souza_loc_prb}.

Following these advances, 
Sgiarovello and co-workers~\cite{sgiarovello_loc_prb},
have computed
the localization lengths for different cubic semiconductors, 
in the framework of the Kohn-Sham
density functional theory~\cite{hk,ks} (DFT).
They showed that the degree of electron localization is quite different
for the various investigated materials. 
These results encourage to pursue the study to other
insulating crystals.

The localization tensor $\langle r_{\alpha} r_{\beta} \rangle_c$,
can be computed from the periodic part of the Bloch
functions $u_{nk}(\textbf{r})$ and their first derivatives with respect
to their wavevector~:
\begin{eqnarray} \label{eq_tens1}
\langle r_{\alpha} r_{\beta} \rangle_c & = &
\frac{V_c}{N (2 \pi)^3}
\int_{BZ} d\textbf{k}
\sum_{n=1}^{N}
\left \{
\left \langle
\frac{\partial u_{n \textbf{k}}}{\partial k_{\alpha}} \right |
\left .
\frac{\partial u_{n \textbf{k}}}{\partial k_{\beta}}
\right \rangle  \right .\nonumber \\
& & \left . - \sum_{n'=1}^{N}
\left .
\left \langle
\frac{\partial u_{n \textbf{k}}}{\partial k_{\alpha}} \right |
u_{n' \textbf{k}}
\right \rangle
\left \langle u_{n' \textbf{k}}
\left |
\frac{\partial u_{n \textbf{k}}}{\partial k_{\beta}}
\right \rangle
\right .
\right \}
\end{eqnarray}
where $V_c$ is the volume of the primitive unit cell, $N$ the number
of doubly occupied bands and $\alpha$, $\beta$ are two cartesian
directions. 
The localization tensor is a {\it global} quantity that
characterizes the occupied Kohn-Sham manifold as a {\it whole}
(all k-points and all bands).
This statement calls for two comments. First, applications
of DFT to solids often make use of the frozen-core and pseudopotential
approximations, while Eq.(\ref{eq_tens1}) requires
an all-electron calculation. Second, the
behavior of core and valence electrons is treated globally while
both kinds of electrons are expected to exhibit strongly different
localization properties interesting to identify
independently.

The localization tensor has been shown~\cite{sgiarovello_loc_prb}
to give a lower bound for the spread of maximally localized
Wannier functions (WF) as defined by Marzari and
Vanderbilt~\cite{marzari_wannier,marzari_aip} (hereafter 
cited as MV). In order
to get some insight into the physics of the chemical bonds in
molecules and solids, such WF are usually constructed considering
only a restricted number of electronic bands close to the Fermi level.
The spread of the resulting WF is strongly dependent of the
electronic states included in the minimization process. In this
context, it seems interesting to try to identify the intrinsic
localization of the electrons in a specific set of bands and to
understand how this quantity is affected when including
other bands. This would allow to solve the problem associated to
the use of pseudopotentials and to characterize separately the
behavior of core and valence electrons.

This paper is organized as follows. In Sec. \ref{sec_bbb},
we propose a decomposition of the localization tensor into contributions
originating from isolated sets of bands composing the energy spectrum
of a solid. Using a simple model, we then illustrate the role of the
covalent interactions on the different terms of the
decomposition. We also make a connection between the localization
tensor and the Born effective charges and we discuss the relation
between pseudopotential and all-electron calculations. 
In Sec. \ref{sec_method}, we give the technical details underlying
our first-principles calculations and we point out the differences
between our method and that applied in 
Ref.~\onlinecite{sgiarovello_loc_prb}.
In Sec. \ref{sec_res} and \ref{sec_disc}, we present
the results obtained on two ferroelectric perovskites (BaTiO$_3$
and PbTiO$_3$) as well as on two binary oxides (BaO and $\alpha$-PbO).
We investigate the variation of electron localization during 
the phase transitions of BaTiO$_3$ and show that the evolution is compatible
with the electronic structure of this compound.

\section{Band by band decomposition of the localization tensor} \label{sec_bbb}

\subsection{Formalism}

Contrary to the polarization and the Born effective charges,
for which band-by-band decompositions have been previously
reported~\cite{dcharges2,posternak_zstar,dcharges1,ghosez_bbb}, the
localization tensor (Eq. (\ref{eq_tens1})) involves scalar
products between Bloch functions of different bands, making the
identification of the contribution of isolated sets of bands less
straightforward. In order to explain this fact,
we have to remember that the localization tensor is related to the
second moment of WF while the Born effective charges and the
spontaneous polarization are linked to their first moment. From standard
statistics, it is well known that these two quantities do not add
the same way~: when considering two random variables $x_1$ and
$x_2$, the mean value of the sum $x_1+x_2$ is simply the sum of the
mean values while the variance of the sum is the sum of the variances
{\it plus} an additional term, the covariance.

These considerations can be transposed in the simple context of a
confined model system made of two orthonormalized states $\psi_1(x)$
and $\psi_2(x)$. The total many-body wavefunction $\Psi(x_1,x_2)$ is a Slater
determinant constructed on the one-particle orbitals.
The center of mass  is given by the expectation
value of the position operator $\hat{X} = \sum_{i=1,2} \hat{x}_i$
\begin{equation}
\overline{X}  = 
\langle \Psi | \hat{X} | \Psi \rangle =
\sum_{i=1,2} \langle \psi_i | \hat{x} | \psi_i \rangle  
\end{equation}
while the total spread (two times the localization tensor)
is related to $\hat{X}^{2}$,
\begin{eqnarray}
\sigma^2 & = &
\langle \Psi | \hat{X}^2 | \Psi \rangle - 
\langle \Psi | \hat{X} | \Psi \rangle^2
\nonumber \\
&=&
\sum_{i=1,2} [\langle \psi_i | \hat{x}^2 | \psi_i \rangle - \langle \psi_i 
| \hat{x} | \psi_i \rangle^2 ]
-
2 \langle \psi_1 | \hat{x} | \psi_2 \rangle 
\langle \psi_2 | \hat{x} | \psi_1 \rangle.
\end{eqnarray}
We see that the first moments of the one-particle orbitals add to form
the total dipole of the many-body wavefunction.
On the contrary, the total spread 
is not equal to the sum of the individual
spreads of $\psi_1$ and $\psi_2$ but involves also matrix elements
of the one-particle position operator $\hat{x}$ between $\psi_1$ and $\psi_2$.
The additional term would be absent if the many-body wavefunction
was a simple product of the one-particle orbitals. It arises
from the anti-symmetry requirement. In analogy with the
language of statistics, we will name it the {\it covariance}.


Based on these arguments, we can now define a band-by-band decomposition
of Eq. (\ref{eq_tens1}). Suppose that the band structure is formed of
$N_g$ groups labelled $G_i$, each of them composed 
of $n_i$ bands ($i = 1, ..., N_g$). The
variance of a particular group $G_i$ is defined as
\begin{eqnarray} \label{eq_bbb4}
\langle r_{\alpha} r_{\beta} \rangle_c(G_i) & = &
\frac{V_c}{n_i (2 \pi)^3}
\int_{BZ} d\textbf{k}
\left \{
\sum_{n \in G_i}
\left \langle
\frac{\partial u_{n \textbf{k}}}{\partial k_{\alpha}} \right |
\left .
\frac{\partial u_{n \textbf{k}}}{\partial k_{\beta}}
\right \rangle  \right .\nonumber \\
& & \left . - \sum_{n,n' \in G_i}
\left .
\left \langle
\frac{\partial u_{n \textbf{k}}}{\partial k_{\alpha}} \right |
u_{n' \textbf{k}}
\right \rangle
\left \langle u_{n' \textbf{k}}
\left |
\frac{\partial u_{n \textbf{k}}}{\partial k_{\beta}}
\right \rangle
\right .
\right \}
\end{eqnarray}
where the sums have to be taken over the bands of group $G_i$.
The covariance of two groups $G_i$ and $G_j$ ($i \neq j$) 
is given by the following relationship:
\begin{equation}\label{eq_bbb5}
\langle r_{\alpha} r_{\beta} \rangle_c(G_i,G_j) =
\frac{-V_c}{n_i n_j (2 \pi)^3}
\int_{BZ} d\textbf{k}
\sum_{n \in G_i} \sum_{n' \in G_j}
\left .
\left \langle
\frac{\partial u_{n \textbf{k}}}{\partial k_{\alpha}} \right |
u_{n' \textbf{k}}
\right \rangle
\left \langle u_{n' \textbf{k}}
\left |
\frac{\partial u_{n \textbf{k}}}{\partial k_{\beta}}
\right \rangle.
\right .
\end{equation}
Using these definitions, the total tensor, associated to the whole
set of occupied bands, can be written as
\begin{equation} \label{eq_bbb6}
\langle r_{\alpha} r_{\beta} \rangle_c =
\frac{1}{N}
\sum_{i=1}^{N_g} n_i
\left \{
\langle r_{\alpha} r_{\beta} \rangle_c(G_i) +
\sum_{j \neq i}^{N_g} n_j
\langle r_{\alpha} r_{\beta} \rangle_c(G_i,G_j)
\right \}.
\end{equation}

The variance 
$\langle r_{\alpha} r_{\beta} \rangle_c(G_i)$
is intrinsic to an isolated set of bands. 
It is related~\cite{souza_loc_prb,sgiarovello_loc_prb}
to the quantity $\Omega_I$ introduced by MV through the relation 
\begin{equation}
\Omega_I = n_i \sum_{\alpha=1}^{3}\langle r_{\alpha} r_{\alpha} \rangle_c(G_i).
\end{equation}
In a one-dimensional crystal, $\Omega_I$ is simply 
the lower bound of the total spread $\Omega$ of the WF~\footnote{
$\langle ... \rangle_n$ represents the expectation value over the
$n^{th}$ occupied Wannier function in the unit cell},
\begin{equation} \label{eq_mvspread}
\Omega = \sum_{n \in G_i} [ 
\langle r^2 \rangle_n -
\langle r \rangle_n^2 ], 
\end{equation}
that can be realized by choosing an adequate phase factor for the Bloch
functions.
In a three-dimensional crystal, it is no more possible
to construct WF that are simultaneously maximally localized in all
cartesian directions. It is only possible to minimize
their spread in one given direction as realized for
the so-called hermaphrodite orbitals introduced in 
Ref.~\onlinecite{sgiarovello_loc_prb}~: these 
particular functions are localized
(Wannier-like) in a given direction $\alpha$ and delocalized
(Bloch-like) in the two others. 
The variance of a particular group of bands
$\langle r_{\alpha} r_{\alpha} \rangle_c (G_i)$ is the 
lower bound of the {\it average} spread 
$
\frac{1}{n_i} \sum_{n \in G_i} [ 
\langle r_{\alpha}^2 \rangle_n -
\langle r_{\alpha} \rangle_n^2 ] 
$
where the sum is taken
over all Wannier-like functions in the unit cell belonging
to group $G_i$. This lower bound is reached for WF that are maximally
localized in direction $\alpha$.
The variance therefore gives some insight
on the localization of the electrons within a specific set of bands taken
independently. This localization is affected by the hybridizations
between atomic orbitals giving rise to the formation of the
considered electronic bands within the solid so that the variance can act
as a probe to characterize these hybridizations.

The covariance is no more related to an isolated set of bands.
It teaches us how the construction of WF
including other bands can improve the localization. 
As discussed by MV, the definition of groups of bands in a solid is 
not unique and sometimes
there is a doubt about which bands have to be considered together.
If we consider two sets of bands $G_i$ and $G_j$ as one single group,
its total variance is the sum of the individual 
variances {\it and} covariances,
that have to be rescaled by the number of bands in each group
\begin{equation} \label{eq_bbb7}
\langle r_{\alpha} r_{\beta} \rangle_c =
\frac{1}{n_i + n_j}
\left \{
n_i \left [
\langle r_{\alpha} r_{\beta} \rangle_c(G_i) +
n_j \langle r_{\alpha} r_{\beta} \rangle_c(G_i,G_j)
\right ] +
n_j \left [
\langle r_{\alpha} r_{\beta} \rangle_c(G_j) +
n_i \langle r_{\alpha} r_{\beta} \rangle_c(G_j,G_i)
\right ]
\right \}.
\end{equation}

Until now, we considered separately the two cartesian directions
$\alpha$ and $\beta$. Stronger results can be obtained
when diagonal elements of the localisation tensor are
considered, or when this localication tensor is diagonalized,
and the eigenvalues are considered. Different inequalities
can be derived. In particular,
from Eq. (\ref{eq_bbb5}), it appears that the covariances
for $\alpha=\beta$ are always {\it negative}. 
This means that 
the diagonal elements of the full tensor are always smaller than 
those obtained by the
sum of the diagonal variances. 
In other words, it is always possible to obtain more
strongly localized orbitals by constructing WF considering
more than one group of bands.
As a consequence
the covariance appears as a tool to identify which bands have to be
considered together in the construction of WF in order to
improve their localization.

In appendix \ref{sec_opticalc} we give an interpretation
of the variance and covariance in terms of the optical conductivity.
It illustrates from a different viewpoint the influence of
the fermionic nature of the electrons on the localization tensor~:
the appearance of the covariance in Eq. (\ref{eq_bbb6}) 
is a direct consequence of the Pauli principle.

\subsection{Simple model} \label{sec_modeldiat}

In this section we will investigate 
a one-dimensional model system. This will help us to understand
the role of the covalent interactions on the electron localization
length
and related quantities like the Born effective charges.
We will deal with a confined system for which
the localization tensor can be computed from matrix elements
of the position operator and its square as described
in Ref. \onlinecite{sgiarovello_loc_prb}. 

Let us consider a diatomic molecule XY. In order to describe the chemical
bonds of this model system we adopt a tight-binding scheme~\cite{harrison}
defined by the hopping integral $t$ and the on-site terms 
$\Delta$ and $- \Delta$. We will call $a$ the interatomic distance
and $\psi_X$, $\psi_Y$ the s-like atomic orbitals that are used as basis
functions. 
The hamiltonian can be rescaled by $\Delta$ (A=t/$\Delta$) 
in order to become
a one parameter hamiltonian defined by
\begin{equation} \label{eq_model1}
H = \left (
\begin{array}{cc}
-1 & A \\
A & 1 \\
\end{array}
\right ).
\end{equation}
We further assume that $\psi_X$ is centered at the origin,
$\psi_Y$ in $a$ and that these two functions do not overlap
at any $x$
\begin{equation} \label{eq_model2}
\psi_X(x) \psi_Y(x-a) = 0.
\end{equation}
The eigenfunctions of the hamiltonian correspond to
\begin{equation} \label{eq_model3}
\phi_{1,2}(x) = 
u_{1,2} \psi_X(x) + v_{1,2} \psi_Y(x-a)
\end{equation}
where the coefficients $u_{1,2}$ and $v_{1,2}$ can be expressed
in terms of the bond polarity~\cite{harrison} 
$\alpha_p$ ($\alpha_p = \frac{1}{\sqrt{1+A^2}}$):
\begin{equation}
\begin{array}{ll}
u_{1} = \sqrt{\frac{1 + \alpha_p}{2}}, &
v_{1} = \sqrt{\frac{1 - \alpha_p}{2}} \\
u_{2} = \sqrt{\frac{1 - \alpha_p}{2}}, &
v_{2} = -\sqrt{\frac{1 + \alpha_p}{2}}. \\
\end{array}
\end{equation}

In order to see the meaning of the different terms appearing in the
band by band decomposition of the localization tensor and the
Born effective charges let us first consider the molecular orbitals 
independently. 

The variance of state $\phi_1$ can be computed from the coefficients
$u_1$ and $v_1$. It writes
\begin{equation} \label{eq_model_var}
\langle x^2 \rangle_c (1) =
\sigma_X^2 \frac{1+\alpha_p}{2} +
\sigma_Y^2 \frac{1-\alpha_p}{2} +
\frac{a^2 A^2}{4 (1+A^2)}
\end{equation}
where $\sigma_X^2$ and $\sigma_Y^2$ are the second central moments
of $\psi_X$ and $\psi_Y$. The variance of $\phi_2$ is given by
a similar expression.
This quantity is composed of three positive terms that summarize the 
mechanisms that are able to delocalize the electrons 
with respect to the atomic orbitals. 
On one hand, the electronic
cloud on a particular atom is not a delta-Dirac function but presents a
degree of delocalization related to $\sigma_X^2$ and $\sigma_Y^2$
(first and second term). When the state $\phi_1$ is made entirely
of $\psi_X$, that is, when $\alpha_p$ equals one, the localization
length is correctly equal to $\sigma_X^2$ (first term). Incorporating
more $\psi_Y$ changes the localization length in proportion 
of $\alpha_p$ (the balance between first and second terms).
On the other hand, the
electrons can occupy two sites $X$ and $Y$ that are separated
by a distance $a$ (third term). This term scales as
$a^2$. Even a small covalent interaction is thus
able to induce an important delocalization if it acts on a sufficiently
large distance. 

The Born effective charge of atom $X$ is defined as the derivative
of the dipole moment $p$ with respect to $a$.
This dipole moment is the sum of the nuclear and static electronic
charges multiplied by the interatomic distance. The contribution coming
from the electrons occupying state $\phi_1$ is equal to
\begin{equation} \label{eq_dip}
p_1 = -2 e u_1^2 a = - e (1+\alpha_p)a
\end{equation}
where $e$ is the module of the electronic charge.
The derivative of Eq. (\ref{eq_dip}) with respect to $a$ gives the
contribution of these electrons to the total effective charge
\begin{equation} \label{eq_zast}
Z^{\ast}_{X,1} = \frac{\partial p_1}{\partial a}
= -e(1+\alpha_p) + e a \frac{A}{(1+A^2)^{3/2}} \frac{\partial A}{\partial a}.
\end{equation}
The first term is the (static) effective atomic charge~\cite{harrison}
of atom X while the second term represents an additional dynamical
contribution due to a transfer on electrons between $X$ and $Y$ during
a relative atomic displacement.
The contribution of the electrons occupying state $\phi_2$
is given by a similar expression
\begin{equation} \label{eq_zast2}
Z^{\ast}_{X,2} = \frac{\partial p_2}{\partial a}
= -e(1-\alpha_p) - e a \frac{A}{(1+A^2)^{3/2}} \frac{\partial A}{\partial a}.
\end{equation}

This simple model illustrates how both the variance 
of the localization tensor and the Born effective
charges depend on the covalent interactions defined by the parameter $A$.
The variance is a {\it static} quantity depending 
on the amplitude of the covalent interations only while the 
the Born effective charges are {\it dynamical} quantities that 
also depend on the variations 
of these interactions during
a relative atomic displacement. 

If we now consider the states $\phi_1$ and $\phi_2$ as a single group 
we have to add their variances and covariances to get the whole
localization tensor. The covariance reduces to
\begin{equation} \label{eq_cov}
\langle x^2 \rangle_c (1,2) =
\frac{-a^2 A^2}{4(1+A^2)}. 
\end{equation}
By adding this covariance to the variance in Eq. (\ref{eq_model_var}), we
remove in some sense the delocalization induced by the covalent
interactions. The total localization tensor becomes independent of the
hopping $A$ and the interatomic distance $a$. It reduces to the mean
spread of the atomic orbitals $\psi_X$ and $\psi_Y$~:
\begin{equation} \label{eq_model_loctens}
\langle x^2 \rangle_c =
\frac{\sigma_X^2+\sigma_Y^2}{2}.
\end{equation}
Eq. (\ref{eq_model_loctens}) defines the mean spread of the  WF constructed
as linear combinations of $\phi_1$ and $\phi_2$ that minimize
the spread functional $\Omega$ (see Eq. (\ref{eq_mvspread})).
As shown by MV, these WF diagonalize the position
operator $\hat{x}$ projected on the subspace of occupied states.
They are thus equal to the atomic orbitals since 
the hypothesis of zero overlap (Eq. (\ref{eq_model2})) implies
$ \langle \psi_X | \hat{x} | \psi_Y \rangle = 0$.

The total Born effective charge of atom $X$ 
can be obtained by adding the nuclear charge $Z^{\ast}_{core}=2 e$
to the terms (\ref{eq_zast}) and (\ref{eq_zast2}).
It is easy to check that for this model 
$Z^{\ast}_{X}$ is equal to zero. This result
can be interpreted in two ways. The point of
view usually adopted is to say that the two molecular orbitals are of the
opposite polarity so that the total dipole of the molecule vanishes.
Based on the results of the preceeding paragraph, we can
also affirm that each maximally localized WF is confined on a single
atom so that no interatomic charge transfer can take place.

This result suggests that the variance gives more
informations about the localization of electrons of
particular chemical bonds than the total localization tensor.
It also illustrates the observation of Ghosez 
{\it et al.}~\cite{dcharges1,dcharges2} that anomalous effective
charges mainly come from hybridizations between occupied and unoccupied
states. In fact, the different chemical bonds generate opposite effects
so that a net charge transfer is possible only if some of them
are unoccupied.

In summary, we have illustrated the mechanisms that govern the variance
of the localization tensor
and the Born effective charges in the particular case of a 
one dimensional model system.
The observations made in this section give us an intuitive understanding
of how delocalized electrons can generate {\it anomalous} effective
charges. Hybridizations between occupied states generate
opposite effects that tend to cancel out when they are summed.
Because of the simplicity of the above adopted
picture, we have however to be careful when we apply this model
to real materials. First, we considered only hybridizations
between two types of atomic orbitals, while the chemical bonds in
real systems generally result from more complicated interactions.
In particular, we neglected on-site hybridizations that are also
able to generate {\it anomalous} effective
charges but that induce a stronger localization 
on the electronic cloud.
Second, 
the hypothesis of zero
overlap (\ref{eq_model2}) is not always fullfilled so that maximally
localized WF constructed on the whole set of occupied states 
generally not reduce to the atomic orbitals.
Nevertheless, this simple model will allow us to interpret some
results in Sections \ref{sec_res} and \ref{sec_disc}.

\subsection{Pseudopotentials}

As mentioned in the introduction, there is a fundamental
problem in the computation of the total localization tensor 
when pseudopotentials are used. 
This is due to the fact that the localization tensor is related
to the bands of the system as a whole~: first, there is no cancellation
between the core electrons and the nuclear charge, as it is
the case in the
computation of the total polarization; second,
the localization tensor is a kind of mean over all bands,
that combines strongly localized (core) states, and
weakly localized (valence) states.
This is clearly 
seen in Eq.(\ref{eq_tens1}), where the 
number of bands explicitely appears 
both as the
denominator of the prefactor and in the two summations.
The band-by-band decomposition
allows us to overcome this problem partly, by 
focusing only on the variances
of isolated groups of bands. Thanks to Eq. (\ref{eq_bbb7})
it is also possible to get some insight into the physics of the 
all-electron localization tensor when pseudopotentials are used.
In this section, we focus on the diagonal elements of the electron
localization tensor $\alpha=\beta$
(of course, any direction can be chosen as $\alpha$).

In an all-electron calculation, let us consider
separately two sets of bands: core bands 
(labelled as 'co'), and valence bands (labelled as 'va'). 
The total localization tensor
can be obtained from the localization tensors of each
group of bands, combined with the covariance between the
two groups of bands:
\begin{equation} \label{all_electron}
\langle r_{\alpha} r_{\alpha} \rangle_c =
\frac{1}{n_{co} + n_{va}}
\left \{
n_{co} 
\langle r_{\alpha} r_{\alpha} \rangle_c(co) +
n_{va}
\langle r_{\alpha} r_{\alpha} \rangle_c(va) +
2 n_{co} n_{va}
\langle r_{\alpha} r_{\alpha} \rangle_c(co,va)
\right \}.
\end{equation}

Both variances 
$\langle r_{\alpha} r_{\alpha} \rangle_c(co)$
and 
$\langle r_{\alpha} r_{\alpha} \rangle_c(va)$ 
are positive quantities. 
The covariance times the product
of the number of bands
$ n_{co} n_{va}
\langle r_{\alpha} r_{\alpha} \rangle_c(co,va)$, 
a negative quantity,
 must always be smaller
in magnitude than each of the related variances 
multiplied by the corresponding number of bands.
This translates to bounds 
on the diagonal elements of the total localization tensor~:
\begin{equation} \label{pseudos1}
\frac{
\left |
n_{va}
\langle r_{\alpha} r_{\alpha} \rangle_c(va) -
n_{co}
\langle r_{\alpha} r_{\alpha} \rangle_c(co)
\right |
}{n_{co} + n_{va}}
\le
\langle r_{\alpha} r_{\alpha} \rangle_c
\le
\frac{
n_{va}
\langle r_{\alpha} r_{\alpha} \rangle_c(va) +
n_{co}
\langle r_{\alpha} r_{\alpha} \rangle_c(co)
}{n_{co} + n_{va}}.
\end{equation}

In the frozen-core approximation,
$\langle r_{\alpha} r_{\alpha} \rangle_c(co)$ 
can be obtained from
separate all-electron calculations for each atom
of the system. 
The localization tensor of the valence bands 
is (likely) computed accurately 
in the pseudopotential approximation~:
the spread of the Wannier functions should be quite similar
if estimated from all-electron valence wavefunctions
or from pseudo-wavefunctions. 

Thus, a {\it bound} on the diagonal
elements of the localization tensor
{\it can be computed} 
from the atomic wavefunctions of the core electrons
and the pseudo-valence wavefunctions. 
In order to compute
the covariance more accurately it is necessary to reconstruct the 
all-electron wavefunctions. This could be done following the ideas
exposed in Ref. \onlinecite{paw}.

\section{Method and implementation} \label{sec_method}

In the remaining part of this paper, we apply the previous formalism
to various oxides. The electronic wavefunctions are obtained within
DFT~\cite{hk,ks} and the local density approximation (LDA) thanks
to the {\sc abinit}~\cite{abinit} package. At variance with a previous
work on semiconductors~\cite{sgiarovello_loc_prb}, the first derivatives
of the wave functions with respect to their wavevector
are not computed from finite differences but
from a linear-response approach~\cite{xggrad} within the
parallel-transport gauge. The wave functions are further transformed
to the diagonal gauge~\cite{ghosez_bbb}.
Both ground-state and first-order wavefunctions are expanded in plane waves up
to a kinetic-energy cutoff of 45 Hartree. Integrations over the
BZ are replaced by sums over a $8 \times 8 \times 8$ mesh of special
k-points~\cite{mp}. With these parameters, the convergence of the localization
tensor for the investigated compounds is better than
10$^{-3}$ Bohr$^2$.
The ionic-core electron potentials of the Ba, Pb, Ti and O
atoms are replaced by ab initio,
separable, extended norm-conserving pseudopotentials, as proposed
by M. Teter~\cite{teter_ext}.
Ba 5s, 5p and 6s electrons, Pb 6s, 5d and 6p electrons,
Ti 3s, 3p and 3d electrons, O 2s and 2p electrons
are considered as valence states.
Beside calculating the localization tensor on bulk-materials,
we also computed it on the isolated atomic systems
Ba$^{2+}$, Pb$^{2+}$ and O by placing each atom at the origin
of a periodic supercell of 20 Bohrs.

As shown by Sgiarovello {\it et al.}~\cite{sgiarovello_loc_prb},
the localization tensor and thus the variances and covariances, are
real. Moreover, they are obviously symmetric in $\alpha$ and $\beta$.
Consequently there exists a set of cartesian axes where they are 
diagonal and their eigenvalues are also real numbers. In the discussion
of our results we will always work in this particular frame so that
we do not need to consider the off-diagonal elements of the
localization tensor.

\section{Results} \label{sec_res}

\subsection{Structural and electronic properties}

We will consider the two binary oxides BaO and $\alpha$-PbO as well as the
ferroelectric perovskites BaTiO$_3$  
and PbTiO$_3$. 
BaO has a rocksalt structure while the tetragonal $\alpha$ phase of 
lead oxide is formed of parallel layers of Pb and O atoms.
BaTiO$_3$ and PbTiO$_3$ have a high-temperature cubic perovskite structure
with five atoms per unit cell. As the temperature is lowered,
the former compound undergoes a sequence of three ferroelectric
phase transitions transforming to tetragonal, orthorhombic and 
rhombohedral structures while the latter compound only undergoes one
single transition from the cubic to the tetragonal phase. We will
consider explicitely the cubic, tetragonal and rhombohedral
phases of BaTiO$_3$ as well as the cubic phase of PbTiO$_3$.

\begin{figure}[htb]
\begin{center}
\includegraphics[width=8cm]{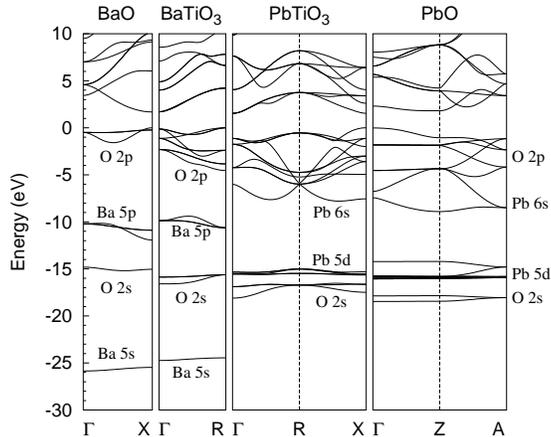}
\end{center}
\caption{\label{fig_bdstr} Band structures of BaO, cubic BaTiO$_3$,
    cubic PbTiO$_3$ and $\alpha$-PbO.}
\end{figure}

The electronic structures of 
these compounds have
been previously 
studied~\cite{posternak_zstar,ghosez_ferroelectrics,pbo_watson,pbtio3a} 
and are illustrated in Fig. \ref{fig_bdstr}. They are formed of well
separated groups of bands. Each of them has a marked dominant orbital
character and can be labeled by the name of the atomic orbital that
mainly composes the energy state in the solid.
The bands at the Fermi level are mainly composed of O 2p states that
show significant interactions with other atomic orbitals like
the well known O 2p-Ti 3d hybridization in BaTiO$_3$ and PbTiO$_{3}$.
The bandstructures in the ferroelectric phases of BaTiO$_3$ are
similar to that of the cubic phase. The phase transitions principally affect
the bandgap and the spread of the O 2p bands while the positions of
the deeper lying bands remain quite constant.
The main difference in the electronic structures of BaO and
BaTiO$_3$ on one hand and PbO and PbTiO$_3$ on the other hand
comes from the presence or absence of Pb 6s electrons 
(that form the so called lone-pair
in PbO). These electrons show a strong hybridization with the O 2p
states. As a consequence the O 2p and Pb 6s bands are degenerate 
at the $R$ point in PbTiO$_3$ and around the $Z$ point in PbO.
Consequently,
we have to consider them as one single group of bands in the
decomposition of the localization tensor.

\subsection{Localization tensor and Born effective charges}

As the total localization tensor is meaningless in pseudopotential
calculations that do not include covariances with the core states,
we focus on the variances of the different groups of bands.
The values can be found in the Tables \ref{tab_ba} and \ref{tab_pb}
where they are compared to the variances of the dominant atomic orbitals.
We do not report any values associated to the deepest lying 
Ti 3s and Ti 3p bands although they have been included in our pseudopotential
calculation. Their variances are in fact close to the atomic ones
and they do not show any sizeable covariance with other bands in
both BaTiO$_3$ and PbTiO$_3$.

\begin{table}[htbp]
\caption{\label{tab_ba} Variances (Bohr$^2$) of the Ba 5s, O 2s, Ba 5p
   and O 2p bands for the isolated atomic systems Ba$^{2+}$ and O,
   BaO and the cubic (C), tetragonal (T) and rhombohedral (R) phases
   of BaTiO$_3$.}
\begin{ruledtabular}
\begin{tabular} {lcccccc}
System   & Str.      & Element & \multicolumn{4}{c}{Band} \\
          & &           & Ba 5s & O 2s  & Ba 5p & O 2p \\
\hline
Atom     & $-$       & $\langle r^2 \rangle_c$
                        & 1.011 & 0.929 & 1.370 & $-$   \\
BaO      & $-$       & $\langle r^2 \rangle_c$
                        & 1.065 & 1.552 & 2.023 & 2.199 \\
BaTiO$_3$&  C        & $\langle r^2 \rangle_c$
                        & 1.091 & 0.950 & 2.189 & 1.875 \\
          &  T        & $\langle r_{\perp}^2 \rangle_c$
                        & 1.091 & 0.945 & 2.180 & 1.852 \\
          &           & $\langle r_{\parallel}^2 \rangle_c$
                        & 1.088 & 0.965 & 2.175 & 1.842 \\
          &  R        & $\langle r_{\perp}^2 \rangle_c$
                        & 1.092 & 0.963 & 2.196 & 1.862 \\
          &           & $\langle r_{\parallel}^2 \rangle_c$
                        & 1.092 & 0.949 & 2.189 & 1.804 \\
\end{tabular}
\end{ruledtabular}
\end{table}

\begin{table}[htbp]
\caption{\label{tab_pb} Variances (Bohr$^2$) of the O 2s, Pb 5d
   and Pb 6s +O 2p bands in
   PbTiO$_3$, $\alpha$-PbO and for the isolated atomic systems
   Pb$^{2+}$ and O.}
\begin{ruledtabular}
\begin{tabular} {lcccc}
System   & Element                            & \multicolumn{3}{c}{Band} \\
          &                                    & O 2s  & Pb 5d &Pb 6s + O 2p\\
\hline
Atom     & $\langle r^2 \rangle_c$            & 0.929 & 0.657 & $-$        \\
PbTiO$_3$& $\langle r^2 \rangle_c$            & 1.874 & 1.490 & 1.749      \\
PbO      & $\langle r_{\perp}^2 \rangle_c$    & 2.234 & 1.142 & 2.178      \\
          & $\langle r_{\parallel}^2 \rangle_c$& 1.724 & 0.990 & 1.968     \\
\end{tabular}
\end{ruledtabular}
\end{table}

In the cubic crystals BaO, BaTiO$_3$ and PbTiO$_3$ as well as in the
atomic systems, the reported tensors are isotropic so that we only
mention their principal values $\langle r^2 \rangle_c$. This is no more
true in the ferroelectric phases of BaTiO$_3$ where a weak anisotropy
can be observed. The tensors have an uniaxial character as the
corresponding dielectric ones: they are diagonal when expressed
in the principal axes and the elements
$\langle r^2_{\perp} \rangle_c$ and $\langle r^2_{\parallel} \rangle_c$
refer to cartesian directions perpendicular and parallel to the
optical axis (that has the direction of the spontaneous polarization).
A much stronger anisotropy is observed in $\alpha$-PbO where the
localization tensor has the same symmetry as in the ferroelectric
phases of BaTiO$_3$. 
Due to its particular structure formed of
atomic Pb-O planes
the electrons of each group of bands are
more delocalized in a direction parallel
($\langle r^2_{\perp} \rangle_c$) to the atomic planes\footnote{
In $\alpha$-PbO, the optical axis is perpendicular to the atomic layers.}
than perpendicular ($\langle r^2_{\parallel} \rangle_c$)
to them. This observation agrees with our intuitive
picture that the covalent interactions between atoms inside a layer
are stronger than between atoms of different layers.

Examining the variances of the different groups of bands
we see that the Ba 5s electrons show a similar degree of
localization both in BaO and BaTiO$_3$ also equivalent to
that of the corresponding atomic
orbital. On the contrary, the O 2s electrons behave differently
in the materials under investigation: in BaTiO$_3$, their variance
is close to the atomic one while they show a significant larger
delocalization in the three other compounds. It is in fact
surprising to see the degree of delocalization of the inner bands
like the O 2s, Ba 5p or Pb 5d bands. In some cases like
BaTiO$_3$, the electrons of these bands
are even more strongly delocalized than those of the bands at the
Fermi level. These results suggest that the corresponding atomic
orbitals are chemically not inert but present non negligible
covalent interactions. 
An interesting observation can be made for the O 2s and Pb 5d
bands in PbTiO$_3$ and $\alpha$-PbO. The delocalization induced
by the covalent interactions that generate these bands tends
to disappear when we consider them as one single group.
In order to compute the variance of the whole O 2s and
Pb 5d bands,
we have to use Eq. (\ref{eq_bbb7}).
As an example
let us consider PbTiO$_3$. The different elements can be summarized in a
matrix where the diagonal elements are the variances (Bohr$^2$) and the
off-diagonal elements the covariances (Bohr$^2$) of the individual
groups
$$
\left (
\begin{array}{rr}
    1.874  &   -0.240   \\
   -0.240  &    1.490   \\
\end{array}
\right ).
$$
The total variance of the (O 2s + Pb 5d) 
group considered as a whole reduces to 0.734 Bohr$^2$.
For $\alpha$-PbO, we obtain similar values of 0.732 Bohr$^2$ for
$\langle r^2_{\perp} \rangle_c$ and 0.701 Bohr$^2$ for
$\langle r^2_{\parallel} \rangle_c$. These values can be compared to the
mean spread of the atomic orbitals
$
\frac{1}{6} (0.929 + 5 \times 0.657)=0.702 \mbox{ Bohr$^2$}.
$

The results presented above suggest that inner orbitals like
O 2s, Ba 5p or Pb 5d are chemically not inert in the materials
under investigation. This observation seems in contradiction with
the conclusions drawn from partial density of states 
analysis~\cite{pbo_watson} that these states are rather inert.
Nevertheless the inspection of the Born effective charges in
BaO or BaTiO$_3$~\cite{posternak_zstar,dcharges2} confirms
our observations that will now be illustrated for $\alpha$-PbO
and PbTiO$_3$. This points out that the global shape of the
bandstructure is less sensitive to the underlying covalent
interactions than the variance of the localization
tensor or the Born effective charges.

In order to investigate the connection between the localization
tensor and the Born effective charges we report in Table \ref{tab_zstar}
the band by band decomposition of $Z^{\ast}_{Pb}$ in PbTiO$_3$
and $\alpha$-PbO. In the perovskite, this tensor is isotropic while
in $\alpha$-PbO it has the same symmetry as the localization tensor.
The contribution of each group of bands has been separated into a
reference nominal value and an {\it anomalous} charge\footnote{
The Born effective charges are in general compared to an isotropic
nominal value that is the charge expected in a purely ionic compound.
All deviations with respect to this reference nominal value
are referred to as {\it anomalous}.}. For $\alpha$-PbO, we observe
the same anisotropy as for the localization tensor: the covalent interactions
inside an atomic layer ($Z^{\ast}_{Pb \perp}$) generate larger
{\it anomalous} contributions than the interactions involving atoms
of different layers ($Z^{\ast}_{Pb \parallel}$). By looking at the
O 2s and Pb 5d bands we see that they generate important anomalous
charges that confirm our observations concerning the variances of
these bands. Interestingly, in both materials
these contributions cancel out when they are summed.
We observe thus the same tendencies for the Born effective
charges and the localization tensor: the effects induced by the
covalent interactions between inner orbitals tend to disappear
when the resulting bands are considered together.

\begin{table}
\caption{\label{tab_zstar}Band by band decomposition of the Born
effective charges (a. u. of charge) in PbTiO$_3$ and $\alpha$-PbO.
The contributions have been separated into a
reference nominal value and an {\it anomalous} charge.}
\begin{ruledtabular}
\begin{tabular} {lrrrrr}
       & &\multicolumn{1}{c}{PbTiO$_3$}& &\multicolumn{2}{c}{$\alpha$-PbO}\\
Band  & &$Z^{\ast}_{Pb}$& &$Z^{\ast}_{Pb \perp}$&$Z^{\ast}_{Pb \parallel}$\\
\hline
Core        & &     14.00 & &     14.00 &     14.00 \\
O 2s        & &  0 + 3.47 & &  0 + 1.89 &  0 + 0.26 \\
Pb 5d       & &-10 - 3.36 & &-10 - 1.80 &-10 - 0.40 \\
Pb 6s + O 2p& & -2 + 1.78 & & -2 + 1.06 & -2 + 0.48 \\
\hline
Tot.        & &  2 + 1.89 & &  2 + 1.15 &  2 + 0.34 \\
\end{tabular}
\end{ruledtabular}
\end{table}

\section{Discussions} \label{sec_disc}

Based on the simple model exposed in Sec. \ref{sec_modeldiat}
we can suggest the following mechanism to
explain the results presented in the preceeding section. 
The atomic orbitals O 2s and Pb 5d (for which the hypothesis of
zero overlap (\ref{eq_model2}) is reasonable)
present weak covalent interactions that generate the corresponding
energy bands in PbTiO$_3$ and $\alpha$-PbO. When we construct
maximally localized WF for each individual group, the resulting
orbitals are delocalized on Pb and O atoms so that during
an atomic displacement an interatomic transfer of charges $-$ generating
{\it anomalous} Born effective charges $-$ is possible. The fact that the
variance of the global (O 2s + Pb 5d) group of bands is close to the
mean spread of the atomic orbitals suggests that the maximally
localized WF constructed on these bands are similar to the
original atomic orbitals. In other words, they are confined on
a single atom. This confinement also suppresses the interatomic
charge transfer so that the anomalous charges disappear.
We can make similar observations for the Ba 5p and O 2s bands
in BaO and BaTiO$_3$, although,
in the latter compound, the cancellation in the Born effective
charges and the variance is not as complete as in
the three remaining ones.
This suggests that in the lead oxides as well as in BaO, the inner
bands Pb 5d and O 2s (resp. Ba 5p and O 2s) mainly result from
hybridizations between two types of atomic orbitals. At the opposite,
in BaTiO$_3$ the Ba 5p and O 2s bands are formed of more than two types
of atomic orbitals.

Looking now at the bands at the Fermi level, we see that their
variance is significantly larger in BaO and $\alpha$-PbO than
in the corresponding perovskites and that it remains nearly constant
in the different phases of BaTiO$_3$. 
This latter observation
seems surprising for two reasons. (i) The LDA bandgap presents drastic
changes when passing from the cubic (1.72 eV) to the rhombohedral
(2.29 eV) phase. This increase suggests a much stronger
localization of the O 2p electrons in the ferroelectric phases.
(ii) The giant Born effective charges observed in the paraelectric
phase~\cite{dcharges1,dcharges2}
imply an important reorganization of the electronic cloud
during an atomic displacement. It appears surprising that this
reorganization has such small effects on the localization tensor.
These small variations are not restricted to BaTiO$_3$ but similar
observations have been made in other ferroelectric compounds
like LiNbO$_3$~\cite{veithen_aip}.

Considering point (i), we note that the correlation between the
bandgap and the localization tensor is not as tight as one
might think. The variance of the O 2p bands for instance is significantly
larger in BaO than in BaTiO$_3$ in spite of the fact that its LDA
bandgap (1.69 eV) is close to the gap in the cubic phase of BaTiO$_3$.

\begin{figure}[htb]
\begin{center}
\includegraphics[width=6cm]{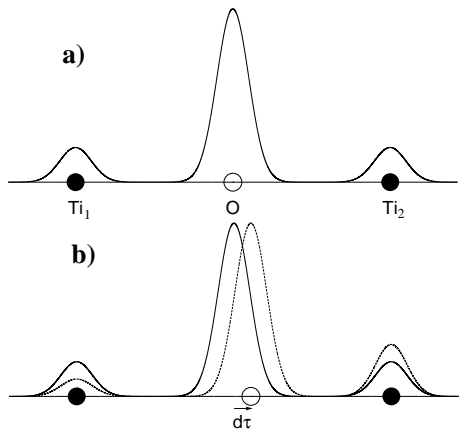}
\end{center}
\caption{\label{fig_wannier} Oxygen centered WF in the cubic phase (solid
line) of
BaTiO$_3$ (a) and its variation during the transition to the
tetragonal phase (dashed line) (b).}
\end{figure}

Considering point (ii), we note that it is possible to have
an important reorganization of the electronic charge {\it without} affecting
the localization tensor a lot. Following the ideas of the Harrison
model~\cite{harrison}, the giant effective charges in perovskite
ferroelectrics result from dynamical orbital hybridizations changes
generating interatomic transfers of charges. In Fig. \ref{fig_wannier} (a)
we have drawn shematically an O centered WF in the cubic phase
of BaTiO$_3$ along a Ti - O chain. 
Due to the O 2p - Ti 3d hybridization, this WF has a
finite probability on the neighbouring Ti$_1$ and Ti$_2$ atoms.
According to the Harrison model, a fraction of electrons is transferred
from Ti$_1$ to Ti$_2$ during a displacement $d\tau$ 
of the O atom (Fig. \ref{fig_wannier} (b)).
Even if the quantity of charges
involved in this process is small, the large scale on which this transfer
takes place (of the order of the lattice parameter) implies a shift
of the WF center larger than the underlying atomic displacement
and explains the {\it anomalous effective charges}.
During the transition from the cubic to the tetragonal phase, the central
O atom is displaced by few percents of the lattice constant $a$ 
($\frac{d\tau}{a} = 0.045$) with respect to Ti$_1$ and Ti$_2$.
The resulting shift of the WF center generates the spontaneous polarization
in the ferroelectric phase.

Based on this simple picture the origin of the small variations of the O 2p
variance during the phase transitions becomes more obvious: 
When the electrons are transferred
from Ti$_1$ to Ti$_2$ their distance to the initial WF center remains
unaffected and their distance to the displaced WF center slightly
decreases due to its shift towards Ti$_2$. Mathematically speaking,
due to the fact that the variations do not depend on the direction
of the atomic displacement,
they are of the second order in
$\frac{d\tau}{a}$.

In order to get a numerical estimate of the charges transferred during
this process and its impact on the localization tensor we can consider
a one dimensional model WF whose square is the sum of three delta-Dirac
functions
\begin{equation} \label{eq_model}
\left | W_n(x) \right |^2 =
\frac{1}{2}
\left \{
\frac{2-Z'_O}{2} \left [\delta(x-a) + \delta(x+a) \right ]
+ Z'_O \delta(x)
\right \}.
\end{equation}
This model only takes into account the delocalization of the electrons
on different atoms (third term of Eq. (\ref{eq_model_var})) while it completely
neglects the delocalization of the electronic cloud on the individual
atoms (first and second term of Eq. (\ref{eq_model_var})).
In this particular case we can identify the localization tensor to the
second moment of the WF. This is no more true in a real, three dimensional
crystal. In BaTiO$_3$ for instance, the O 2p group contains 9 different
WF per unit cell located on three different O atoms. These WF extend in
different spatial directions so that their 
{\it average} spread in the $x$-direction
is lower than the spread of one single WF
as the one shown in Fig. \ref{fig_wannier}.

In Eq. (\ref{eq_model}),
$Z'_O$ represents the probability of the electrons to be found on
the O atom. It can be computed from the value of the O 2p variance
in the paraelectric phase of BaTiO$_3$ and the lattice constant $a$
using the relation
$
\int x^2 |W_n(x)|^2 dx = \langle r^2 \rangle_{c,O2p}.
$
This yields
$Z'_O = 1.73$. This quantity allows an estimate of the static charge
of the O atom in BaTiO$_3$ by substracting three times $Z'_O$ from
the charge due to the nucleus and the core electrons O 1s and O 2s.
This yields $Z_{O,st} = 4 - 3 \cdot 1.73 = -1.19~e$.

When the O atom is displaced,
the shift of the WF center is directly related to
the quantity of charges $\varepsilon$ transferred from Ti$_1$ to Ti$_2$. 
The value of $\varepsilon$ can
be computed from the value of the effective charge 
generated by the O 2p electrons ($Z^{\ast}_{O 2p} = -9.31$) in the
cubic phase~\cite{dcharges1} by taking into account
that the {\it anomalous} charges are generated by three WF located on
the same O atom~\cite{marzari_aip}.  
To get the polarization due to one single WF, we have to divide
this quantity by 3 since each of them brings a similar contribution to
$Z^{\ast}_{O 2p}$. 
In the tetragonal phase, the model WF writes
\begin{equation} \label{eq_wf_tet}
|W_n(x)|^2 = \frac{1}{2} \left \{
\frac{2-Z'_O-\varepsilon}{2} \delta(x+a) + Z'_O \delta(x - d \tau) +
\frac{2-Z'_O+\varepsilon}{2} \delta(x-a)
\right \}.
\end{equation}
By identifying twice its first moment to 
$Z^{\ast}_{O 2p} d\tau/3$ 
one gets
$\varepsilon=0.0614$ at the transition from the
cubic to the tetragonal state. It implies a decrease in the spread of the model
WF of 0.18 Bohr$^2$. 

This variation is larger than the observed one (0.023 Bohr$^2$).
Part of the discrepancy is probably due to the 
fact that we considered $Z^{\ast}_{O 2p}$
to be constant along the path of atomic displacement from the 
paraelectric to the ferroelectric phase. Using the value of 
$Z^{\ast}_{O 2p}$ in the tetragonal phase we obtain a value of 
0.0467 for $\varepsilon$ while the variance decreases of 0.12 Bohr$^2$.
Moreover, one has to bear in mind that the localization tensor
in BaTiO$_3$ is an average value that has to be taken over 9 WF. Six of
them are centered on O atoms that ly in a plane perpendicular
to the direction of the spontaneous polarization. They are
probably less affected by the phase transition. As a consequence,
the variation of the WF located on the remaining O atom (the one
represented on Fig. \ref{fig_wannier}) is expected to be larger than the
variation of the localization tensor.

In summary,
even if there is no formal connection between the real WF in BaTiO$_3$
and Eq. (\ref{eq_model}), this simple model shows that small variations
of the localization tensor are compatible with giant effective
charges and their interpretation in terms of the Harrison model.
As illustrated with the model WF, the transfer of charges along the
Ti$-$O chains only implies a slight decrease in the spread of one single
WF. This {\it decrease} is expected to be larger than the decrease in the
variance because this latter quantity is an
average value over 9 WF that are not modified to the same extent
during the phase transition.

\section{Conclusions}

Using a plane wave-pseudopotential approach to DFT we computed the
electron localization tensor for various oxides. Our study was based
on the work on semiconductors performed by Sgiarovello and
co-workers but used linear-response techniques to compute the
first-order wavefunctions.

In order to investigate the properties of electrons occupying
individual groups of bands independently, we first set-up a band by
band decomposition of the localization tensor. In analogy with the
field of statistics we had to distinguish between variance and
covariance in this decompositon. The significance of these
new concepts was illustrated
in terms of WF and explained on a simple model. The variance 
allows to get some
insight into the hybridizations of atomic orbitals. The covariance can
be useful to help constructing 
maximally localized WF: It identifies the bands that have to be
considered together in order to improve their localization. We also
made a connection between the localization tensor and the Born
effective charges and we discussed the difference between all-electron
and pseudopotential calculations.

We applied these techniques to binary oxides (BaO and $\alpha$-PbO)
as well as perovskite ferroelectrics (BaTiO$_3$ and PbTiO$_3$). By
considering first the electrons of the inner bands we showed that some
of them present a strong delocalization with respect to the
situation in an isolated atom. This observation suggests that the
underlying atomic orbitals are chemically not inert but present non
negligible covalent interactions. This fact had been confirmed from an
inspection of the Born effective charges.

Finally, the variations of the O 2p variance during the ferroelectric
phase transitions of BaTiO$_3$ were found to be very small. This
surprizing result was explained in terms of the electronic
structure of this compound as it is interpreted in the Harrison model.

We think that, when combined with Born effective charges,
the band-by-band decomposition of the localization tensor
could provide a powerful tool for the qualitative characterization
of bonds in solids. However, more studies are needed, for different
classes of materials, in order to make it fully effective.

\section*{Acknowledgments}
M. V. and X. G. are grateful to the National Fund for
Scientific Research (FNRS-Belgium) for financial support. Ph. G.
acknowledges support from FNRS-Belgium (grant 9.4539.00) and the
Universit\'e de Li\`ege (Impulsion grant). This work was supported by the
Volkswagen-Stiftung
(www.volkswagenstiftung.de) within the program "Complex Materials:
Cooperative Projects of the Natural, Engineering, and Biosciences" with the
title: "Nano-sized ferroelectric Hybrids"
under project number I/77 737.
It was also supported by the Communaut\'e Fran\c{c}aise 
through the "Action de Recherche Concert\'ee: 
Interaction Electron-Vibration dans les nanostructures", 
and the Belgian Federal Government,
through the PAI/IUAP P5 "Quantum Phase Effects in Nanostructured Materials"

\appendix

\section{Optical conductivity} \label{sec_opticalc}
The optical conductivity (imaginary part of the optical dielectric
tensor) of a given material is related to its absorption coefficient, 
the probability of the valence electrons to perform optical transitions
to the {\it unoccupied} conduction bands under the influence
of an electromagnetic field.
If we consider only "vertical" band-to-band transitions 
(thus neglecting elementary excitations like the electron-hole interaction
or the electron-phonon coupling) this quantity writes in the 
dipolar approximation~\cite{bassani}
\begin{equation} \label{eq_bbb9}
\varepsilon''_{\alpha \beta}(\omega) =
\frac{4 \pi^2 e^2}{m^2 \omega^2 \hbar}
\sum_{n=1}^{N} \sum_{m=N+1}^{\infty}
\int_{BZ} \frac{2 d\textbf{k}}{(2 \pi)^3}
p_{nm}^{\alpha}(\textbf{k}) p_{mn}^{\beta}(\textbf{k})
\delta \left (
\omega_{mn}(\textbf{k}) - \omega
\right )
\end{equation}
where \textit{m} is the electron mass,
$
\textbf{p}_{nm}(\textbf{k})=-i \hbar 
\langle \psi_{n \textbf{k}} | \nabla \psi_{m \textbf{k}} \rangle
$
and 
$
\hbar \omega_{mn}(\textbf{k}) =
\varepsilon_{m\textbf{k}} - \varepsilon_{n\textbf{k}}
$.
The matrix elements of the momentum operator can equivalently be expressed
as
$
\textbf{p}_{nm}(\textbf{k})=
-m  \omega_{nm}(\textbf{k})
\langle u_{n\textbf{k}} | \partial_{\textbf{k}} u_{m\textbf{k}} \rangle
$.

It has been shown by Souza, Wilkens and Martin~\cite{souza_loc_prb}
that $\varepsilon''$ is related to the localization tensor by the relation
\begin{equation} \label{eq_bbb10}
\int_{0}^{\infty}
\varepsilon''_{\alpha \beta}(\omega) \, d\omega = 
\frac{8 \pi^2 e^2 N}{\hbar V_c}
\left \langle r_{\alpha} r_{\beta} \right \rangle_c.
\end{equation}
In order to see the effect of the band by band decomposition, we will write
$\varepsilon''$ as
\begin{equation} \label{eq_bbb11}
\varepsilon''_{\alpha \beta}(\omega) =
\sum_{i=1}^{N_g} \left \{
\varepsilon''_{\alpha \beta}(\omega;G_i)
+ \sum_{j \neq i}^{N_g}
\varepsilon''_{\alpha \beta}(\omega;G_i,G_j)
\right \}
\end{equation}
where
\begin{subequations}
\begin{eqnarray} 
\varepsilon''_{\alpha \beta}(\omega;G_i) & = &
\frac{4 \pi^2 e^2}{m^2 \omega^2 \hbar}
\sum_{n \in G_i} 
\sum_{
\begin{array}{c}
\vspace{-1.05cm}
\scriptstyle{m \not \in G_i} \\
\scriptstyle{m=1} 
\end{array}}^{\infty}
\int_{BZ} \frac{2 d\textbf{k}}{(2 \pi)^3}
p_{nm}^{\alpha}(\textbf{k}) p_{mn}^{\beta}(\textbf{k})
\delta \left (
\omega_{mn}(\textbf{k}) - \omega
\right ) \label{eq_bbb12a} \\
\varepsilon''_{\alpha \beta}(\omega;G_i,G_j) & = &
\frac{-4 \pi^2 e^2}{m^2 \omega^2 \hbar}
\sum_{n \in G_i} 
\sum_{m \in G_j} 
\int_{BZ} \frac{2 d\textbf{k}}{(2 \pi)^3}
p_{nm}^{\alpha}(\textbf{k}) p_{mn}^{\beta}(\textbf{k})
\delta \left (
\omega_{mn}(\textbf{k}) - \omega
\right ) \label{eq_bbb12b}.
\end{eqnarray}
\end{subequations}
The first sum of Eq. (\ref{eq_bbb12a}) has to be taken over the bands of
group $G_i$ while the second sum extends over all bands (unoccupied or not)
except those of group $G_i$. In Eq. (\ref{eq_bbb12b}), the two sums extend
over the bands of group $G_i$ and $G_j$.
It is easy to show that $\varepsilon''_{\alpha \beta}(\omega;G_i)$
and $\varepsilon''_{\alpha \beta}(\omega;G_i,G_j)$ are
related
to the variances and covariances 
by the relations
\begin{subequations}\label{eq_bbb12}
\begin{eqnarray}
\int_{0}^{\infty}
\varepsilon''_{\alpha \beta}(\omega;G_i) d\omega   & = &
\frac{8 \pi^2 e^2 n_i}{\hbar V_c}
\left \langle r_{\alpha} r_{\beta} \right \rangle_c (G_i) \\
\int_{0}^{\infty}
\varepsilon''_{\alpha \beta}(\omega;G_i,G_j) d\omega   & = & 
\frac{8 \pi^2 e^2 n_i n_j}{\hbar V_c}
\left \langle r_{\alpha} r_{\beta} \right \rangle_c (G_i,G_j). 
\end{eqnarray}
\end{subequations}
Thanks to these definitions, the physical meaning of the covariance
becomes now obvious: If the total localization tensor was simply
the sum of the variances 
$\left \langle r_{\alpha} r_{\beta} \right \rangle_c (G_i)$, the 
expression of the 
dielectric tensor (\ref{eq_bbb9}) would not only contain transitions
between occupied and unoccupied states, but also transitions
between occupied states themselves. It is by adding the covariances
$\left \langle r_{\alpha} r_{\beta} \right \rangle_c (G_i,G_j)$
that one compensates the effect of these forbidden transitions 
in order to get a physically correct quantity.

\end{document}